\documentclass[prl,a4paper,twocolumn,superscriptaddress,longbibliography,english,showpacs]{revtex4-1} 
\usepackage{babel}
\usepackage{xcolor} 
\usepackage{graphicx} 
\usepackage{amsmath} 
\usepackage{amssymb}
\usepackage[backgroundcolor=green!50!white]{todonotes}
\def\be{\begin{equation}} \def\ee{\end{equation}}
\def\bea{\begin{eqnarray}} \def\eea{\end{eqnarray}}
\definecolor{darkblue}{rgb}{0.1,0.2,0.6} \definecolor{darkred}{rgb}{0.8,0.1,0.2}
\usepackage[colorlinks,citecolor=darkblue,linkcolor=darkred,urlcolor=darkblue]{hyperref}
\usepackage[all]{hypcap} 
\makeatletter \adddialect\l@English\l@english \makeatother
\newcommand{\bra}[1]{\langle\,#1\,|} \newcommand{\ket}[1]{|#1\rangle}

\newcommand{\E}{\mathrm{e}}

  \newcommand{\eg}{\textit{e.g.} }
 \newcommand{\etal}{\textit{et al.} }

\renewcommand{\section}[1]{\noindent \textbf{#1} --- }
\begin{document}
\title{Extended slow dynamical regime close to the many-body localization transition}
\author{David J. Luitz}
\affiliation{Department of Physics and Institute for Condensed Matter Theory, University of Illinois at Urbana-Champaign}
\affiliation{Laboratoire de Physique Th\'eorique, IRSAMC, Universit\'e de Toulouse, {CNRS, 31062 Toulouse, France}} 
\email{dluitz@illinois.edu} 
\author{Nicolas Laflorencie}
\affiliation{Laboratoire de Physique Th\'eorique, IRSAMC, Universit\'e de Toulouse, {CNRS, 31062 Toulouse, France}} 
\email{laflo@irsamc.ups-tlse.fr} 
\author{Fabien Alet} 
\affiliation{Laboratoire de Physique Th\'eorique, IRSAMC, Universit\'e de Toulouse, {CNRS, 31062 Toulouse, France}} 
\email{alet@irsamc.ups-tlse.fr}
\date{November 16, 2015}
\begin{abstract} 
Many-body localization is characterized by a slow logarithmic growth of the entanglement entropy after a global quantum quench while the local memory of an initial density imbalance remains at infinite
time. We investigate how much the proximity of a many-body localized phase can influence the dynamics in the  delocalized ergodic regime where thermalization is expected. Using an exact Krylov space technique, the out-of-equilibrium dynamics of the random-field Heisenberg chain is studied up to $L=28$ sites, starting from an initially
unentangled high-energy product state. Within most of the delocalized phase, we find a sub-ballistic entanglement growth $S(t)\propto t^{1/z}$ with a disorder-dependent exponent $z\ge1$, in contrast with the pure ballistic growth $z=1$ of clean systems. At the same time, anomalous relaxation is also observed for the spin imbalance ${\cal{I}}(t)\propto t^{-\zeta}$ with a continuously varying disorder-dependent exponent $\zeta$, vanishing at the transition. This provides a clear experimental signature for detecting this non-conventional regime.
\end{abstract} 
\maketitle


The many-body localization (MBL) phenomenon has attracted an enormous interest in the last few years (see
Refs.~\cite{nandkishore_many-body_2015,altman_universal_2015} for recent reviews). This is mainly
due to the fundamental issues that MBL raises regarding the foundations of quantum statistical physics,
{\it{e.g.}} the absence of thermalization and a violation of the eigenstate thermalization hypothesis
(ETH)~\cite{deutsch_quantum_1991,srednicki_chaos_1994,rigol_thermalization_2008}, the persistence of
local quantum information at very long time~\cite{pal_many-body_2010} and the slow logarithmic growth of entanglement entropy
with time~\cite{chiara_entanglement_2006,znidaric_many-body_2008,bardarson_unbounded_2012,serbyn_universal_2013,vosk_many-body_2013,andraschko_purification_2014}.
Furthermore, MBL behaves as an emerging integrable system, with an extensive number of local
integrals of
motion~\cite{vosk_many-body_2013,serbyn_local_2013,ros_integrals_2015,chandran_constructing_2015},
and MBL states exhibit low (area-law) entanglement even at high energy~\footnote{MPS, DMRG or RSRG-X
    techniques are therefore efficient tools for this localized
    regime, see Refs.~\onlinecite{yu_finding_2015,khemani_obtaining_2015}.}.  In this context, one of the most studied
    theoretical models is the spin-$\frac{1}{2}$ random-field Heisenberg chain~\cite{znidaric_many-body_2008,pal_many-body_2010,bardarson_unbounded_2012,luca_ergodicity_2013,luitz_many-body_2015,agarwal_anomalous_2015}
\be 
{\cal H}=\sum_{i=1}^{L} \left({\vec{S}}_i\cdot {\vec{S}}_{i+1}-h_i S_i^z\right), \label{eq:Hrf}
\ee 
which lies in the same class as interacting fermionic rings in a disordered
potential~\cite{oganesyan_localization_2007,bauer_area_2013,bar_lev_absence_2015,mondaini_many-body_2015}.
Exact diagonalization studies have clearly identified a MBL
transition~\cite{pal_many-body_2010,luitz_many-body_2015,serbyn_criterion_2015}, and a many body
mobility edge in one dimension~\cite{luitz_many-body_2015,serbyn_criterion_2015}, in contrast with single particle
Anderson localization. However the precise
nature of the transition remains elusive despite tentative finite size scaling analyses,
practically limited to the small range of available system sizes $L\le
22$~\cite{luitz_many-body_2015}. 

Recently, two analytical phenomenological renormalization approaches have been proposed by Vosk
\etal~\cite{vosk_theory_2015} and Potter \etal \cite{potter_universal_2015}
for the dynamical transition MBL --- ETH in one dimension.  Building on different ingredients, both studies nevertheless reached comparable conclusions regarding the critical regime. One interesting common aspect is that slow dynamics is predicted on the {\it delocalized} side of the transition,
interpreted as caused by Griffiths regions~\cite{griffiths_nonanalytic_1969}. Signatures of such anomalously slow dynamics
on the ergodic side of the transition was previously observed numerically for 1D models in
Refs.~\cite{agarwal_anomalous_2015,bar_lev_absence_2015,torres-herrera_dynamics_2015} on small systems $L\le 16$. While
Agarwal {\it et al}.~\cite{agarwal_anomalous_2015} found a transition diffusive -- sub-diffusive
roughly in the middle of the ergodic regime, Bar Lev {\it et al}.~\cite{bar_lev_absence_2015}
concluded for a more extended sub-diffusive phase, although they did not precisely locate the
boundary.
\begin{figure}[t!] \centering 
    \includegraphics[width=\columnwidth,clip]{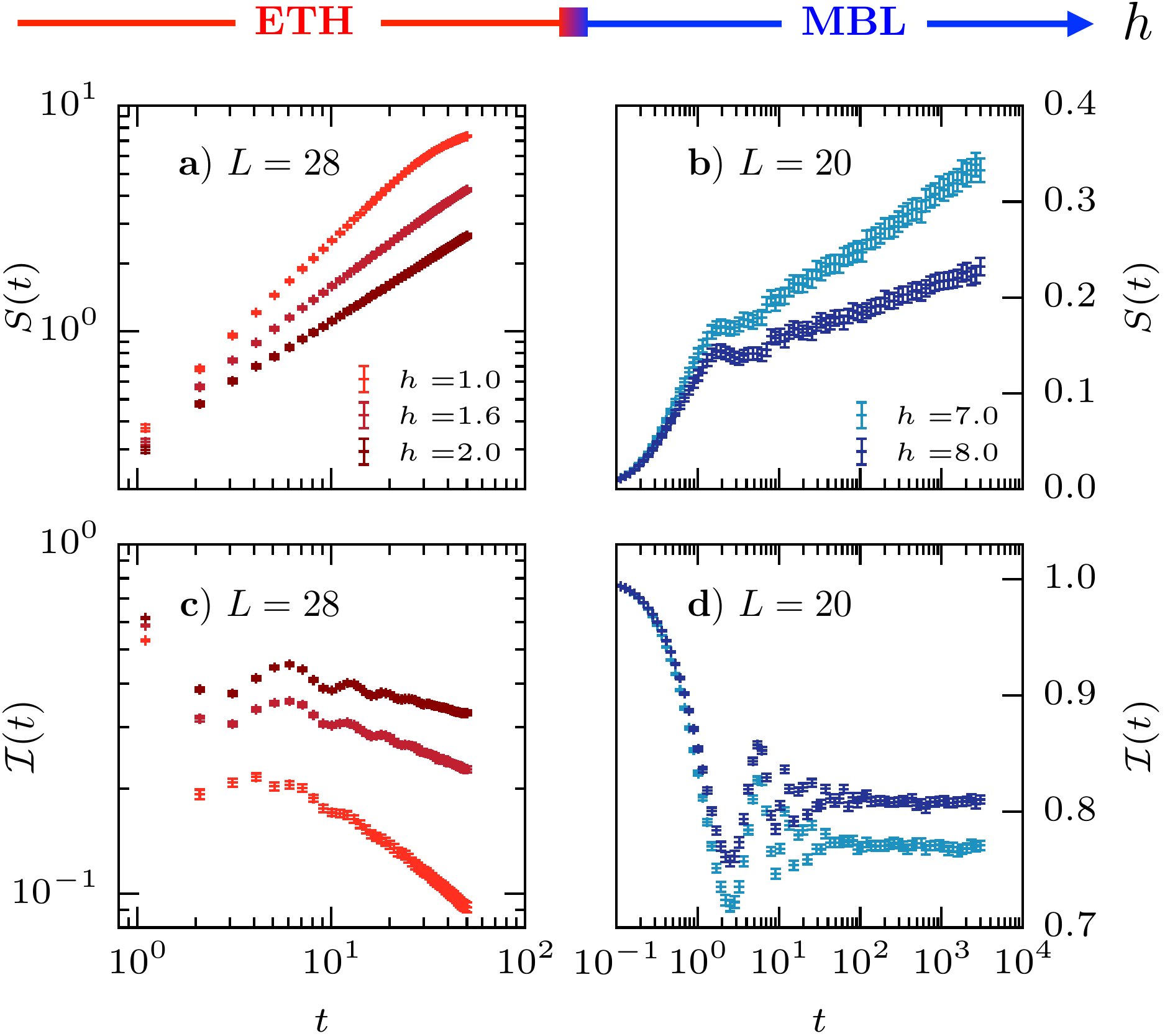}
    \caption{Disorder averaged time evolution of
    the entanglement entropy $S(t)$ [panels a) and b)] for the half-system in an open chain and
    spin density imbalance $\mathcal{I}(t)$ [panels c) and d)], all measured after a quench from a
    random initial product (unentangled) state having an average energy in the middle of the spectrum.
    Left panels show the behavior in the ergodic ETH phase, where the entanglement entropy grows
    as a powerlaw $\propto t^{1/z}$ until saturation and the imbalance decays algebraically $\propto
    t^{-\zeta}$ at intermediate times (ED results for $L=28$ sites). Right panels display the dynamical behavior in the MBL phase, where the entanglement entropy grows logarithmically in time
and the imbalance saturates at a nonzero constant (ED results for $L=20$ sites). Here, we have
averaged over $10^3$ disorder configurations.} \label{fig:overview} \end{figure}

In this Letter we address this crucial issue of anomalous dynamics in the delocalized regime when
approaching the MBL transition for the random-field Heisenberg chain model Eq.~\eqref{eq:Hrf}.  We
study the time evolution after a quantum quench for systems up to $L=28$ sites using an exact Krylov space
method~\cite{nauts_new_1983}. Reaching these large system sizes turns out to be decisive for drawing firm conclusions on the dynamical response after a global quench. We focus on the out-of-equilibrium
response for two key quantities: the entanglement entropy and the spin density imbalance. While the
former is a central object for quantum quenches~\cite{alba_entanglement_2014}, the latter addresses
the prevailing question of how the memory of an initial quantum state is lost with time, and allows
to make a direct connection with recent experiments on interacting fermions in a 1D quasi-random
optical lattice~\cite{schreiber_observation_2015}. 

\begin{figure*}[t]
    \centering
    \includegraphics{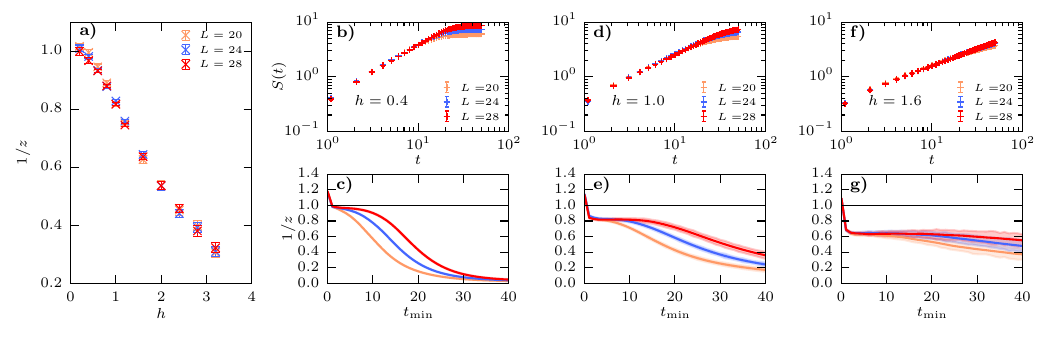}
    \caption{b), d) and f) Disorder averaged time evolution of the entanglement entropy $S(t)$ in the open chain
for different system sizes and three values of disorder. c), e) and g) Logarithmic derivative of
the disorder averaged time evolution of $S(t)$, obtained by power law fits over 8 points in time,
starting from $t_\text{min}$. The formation of plateaus corresponds to the power law regime, with
growing extent in terms of system size. The plateaus determine the range of the power law regime, over
which we extract the exponent $1/z$, displayed as a function of disorder in panel (a). Note that the
range of the power law regime grows with disorder strength as the exponent decreases, delaying the
saturation of $S(t)$. Shaded regions correspond to fit uncertainties.}
    \label{fig:exponent}
\end{figure*}

Our exact numerical results for the time evolution of these two quantities provide a strong support
for the absence of diffusive regime in most of the delocalized ETH phase. Instead, a sub-ballistic
entanglement growth is clearly observed for the von-Neumann entropy $S(t)\propto t^{1/z}$, with a disorder-dependent exponent
$z\geq1$.  The relaxation of an initial spin density imbalance also displays a power-law behavior, as
it decays in time ${\cal{I}}(t)\propto t^{-\zeta}$ with a non-universal exponent $\zeta$, superposed
by sub-dominant oscillatory terms. These two exponents governing the entropy growth and the decay of the
imbalance are continuously varying with the disorder strength and both vanish at the MBL transition.
In the MBL regime, we recover the slow logarithmic growth of
entanglement, while the memory of initial spin density imbalance remains even after long times.
Fig.~\ref{fig:overview} shows an overview of both ETH and MBL regimes for the time evolution of
entanglement and imbalance obtained using Krylov space time evolution with $L=20$ sites in the MBL
regime and $L=28$ in the ETH phase where larger systems are required to capture the slow dynamics.
The power-law regimes with varying exponents can be observed as straight lines in the log-log panels
for the ETH phase.
These exact results (see below for more details) are obtained for initially unentangled product states
filtered such that their energy is in the middle of the many-body spectrum where the critical
disorder strength is $h_c\simeq 3.7$~\cite{luitz_many-body_2015}.\\
\\
{\it Time evolution after a quench---} We consider a global quench protocol, where we follow the time-evolution of an initial product state $\ket{\psi(0)} = \ket{\sigma_1,\dots,\sigma_L}$ given by the $z$ projections $\sigma_i$ under Hamiltonian dynamics
\begin{equation}
    \ket{\psi(t)} = \E^{-i {\cal H} t} \ket{\psi(0)}.
    \label{eq:unitary}
\end{equation}

Studying the dynamics at any arbitrary time by fully diagonalizing ${\cal H}$ is restricted to small system sizes, typically $L=16$ for Eq.~\eqref{eq:Hrf}. Time evolution using variational approaches based on matrix-product states formalism~\cite{vidal_efficient_2003,white_real-time_2004} are particularly successful in cases where the entanglement entropy remains small, \eg in
the MBL phase, but rapidly break down in the ergodic phase due to the fast entanglement growth (see below).
In order to address the ETH regime, we take advantage of the algorithm first proposed in Ref.~\onlinecite{nauts_new_1983} which is based on a
projection of the Hamiltonian to the Krylov space $\mathcal{K}=\text{span}\left(\ket{\psi_0}, {\cal H} \ket{\psi_0}, \dots
{\cal H}^n \ket{\psi_0} \right)$ using the Lanczos algorithm and calculation of the (small) matrix
exponential in the orthonormal Krylov space basis. Here, we use the implementation of the SLEPc
package~\cite{hernandez_slepc:_2005} which calculates the matrix exponential in the Krylov basis by
a simple eigendecomposition. We are able to reach large system sizes for any disorder strength (up
to $L=28$ sites) in the intermediate time regime (up to $t \simeq 10^2$ for the largest systems)
before the entanglement entropy saturates due to finite-system sizes.
As we previously showed~\cite{luitz_many-body_2015} that the critical disorder strength $h_c$ of
the MBL transition depends on the energy of eigenstates, it is 
crucial to specify the energy of the initial state. To this end, we calculate for all
disordered samples the average energy density $\epsilon = \left(\bra{\psi(0)}{\cal
H}\ket{\psi(0)}-E_0\right)/\left( E_1-E_0 \right)$, with $E_0$ ($E_1$) the groundstate (maximal)
energy of the sample, for random basis states $\ket{\psi(0)}$ until we find one whose energy
density is close enough to the desired target density.  In the following, we focus on
initial states with total zero magnetization that are located in the middle of the spectrum
($\epsilon=0.5$). We average our results over at least 1000 disorder realizations, choosing a
different initial state for each sample.\\
\\
{\it Sub-ballistic entanglement growth---} 
We first discuss the time evolution of the entanglement entropy 
\be
S(t)=-{\rm{Tr}}\Bigl[\rho_A(t)\ln\rho_A(t)\Bigr],
\ee
where $\rho_A(t)= {\rm Tr}_B | \Psi(t) \rangle \langle \Psi(t) |$ is the (time-dependent) reduced density matrix obtained after cutting chains of lengths $L=20,24,28$ in two equal parts $A$ and $B$ of size $L/2$.
For clean systems, the growth of entanglement entropy after such a global quench is known to be
ballistic in time~\cite{calabrese_evolution_2005,chiara_entanglement_2006,kim_ballistic_2013}, the
information spreading being limited by a Lieb-Robinson bound~\cite{lieb_finite_1972}. Then, after a
finite time, the entropy will reach its saturation value $S_{\rm sat}=\ell s_\infty$ for a finite
subsystem of length $\ell$ \cite{singh_signatures_2015}, with $s_\infty \simeq \ln 2$ depending on the energy of the initial state (here $s_\infty\simeq \ln 2$ for our initial states with $\epsilon=0.5$).

In practice, the time lapse for observing an asymptotic ballistic regime is restricted to $t<t_{\rm sat}\simeq s_\infty \ell$, which may prevent such an observation in particular for
small system sizes.  Interestingly, using open chains the entanglement entropy
grows a factor of 2 slower as compared to the periodic case, while saturating at the same value $L s_\infty
/2$, thus doubling the time lapse for observing universal entanglement growth before saturation.
The combination of open boundaries and large system sizes is crucial to capture the asymptotic regime for the spreading of entanglement. In the ETH phase of the random-field Heisenberg chain
model Eq.~\eqref{eq:Hrf} at small disorder strength $h$, we see in Fig.~\ref{fig:exponent} a
sub-ballistic growth in time of the entanglement entropy, which follows
\be
\label{eq:Sz}
S(t) \propto t^{1/z},
\ee
with a disorder-dependent dynamical exponent $z\ge1$. The time window over which sub-ballistic
entanglement spreading is visible grows as $(s_\infty L)^z$, which is clearly apparent in
Fig.~\ref{fig:exponent} as plateaus of the local (in time) exponent $1/z$ obtained from sliding fits
to the form Eq.~\ref{eq:Sz} (see caption of Fig.~\ref{fig:exponent}). These local power law fits provide an estimate of how the exponent changes if the fit window is displaced and a plateau indicates a real power law regime. As the observed domains of constant local exponents grow with system size, we conclude that in the thermodynamic limit the entanglement  entropy grows indeed as a power law. For the system sizes $L \leq 16$ accessible to full diagonalization, we find that it is almost impossible to identify such a power-law regime.

The algebraic growth of Eq.~\eqref{eq:Sz} has been predicted to occur in the sub-diffusive regime found in the renormalization approaches of Refs.~\onlinecite{vosk_theory_2015,potter_universal_2015}, with an exponent $z$  which varies continuously with disorder due to the proximity to the critical point. Plotted in panel a) of
Fig.~\ref{fig:exponent}, one sees that $1/z\le 1$ and decreases with $h$. Although it is difficult
to make a definite statement at small disorder strength, it is plausible that the sub-ballistic entanglement spreading regime takes place as soon as $h\ne 0$. In any case this result contrasts with the clearly smaller
sub-diffusive regime found for $L\le 16$  in Ref.~\cite{agarwal_anomalous_2015}.

The exponent $1/z$ is expected to vanish at the ETH-MBL critical point where instead a
logarithmic growth should be
observed~\cite{vosk_theory_2015,potter_universal_2015,luitz_many-body_2015,serbyn_criterion_2015}.
This should also be the case for system sizes below the correlation length in a critical regime
around $h_c$. Within the system sizes and  time regimes that we can access, we cannot discriminate
between a logarithmic and a very slow algebraic behavior. This critical logarithmic growth likely
implies that the power-law fits for $h \gtrsim 3$ may be spoiled by a logarithmic component (not present in our fitting function), resulting in a slightly overestimated value of $1/z$ in this regime.\\
\\
{\it Time evolution of a spin density imbalance---} The hallmark of MBL is the absence of
thermalization, which can be seen in quantum quench protocols as a violation of initial state
independence~\cite{gogolin_absence_2011,nandkishore_many-body_2015}: some memory of the local
initial conditions is preserved even at infinite time, in contrast with the ETH phase where any
particular local feature of the initial state is lost along the unitary evolution. In a recent cold
atom experiment with interacting fermions loaded in a quasi-periodic optical
lattice~\cite{schreiber_observation_2015}, this property has been used to define a working "order
parameter" to characterize the MBL phase for the transition through the study of the relaxation of
an initially prepared charge density wave: a non-zero charge imbalance persisting at long time
signals the MBL phase.

\begin{figure}[t!]
    \centering
    \includegraphics[width=\columnwidth,clip]{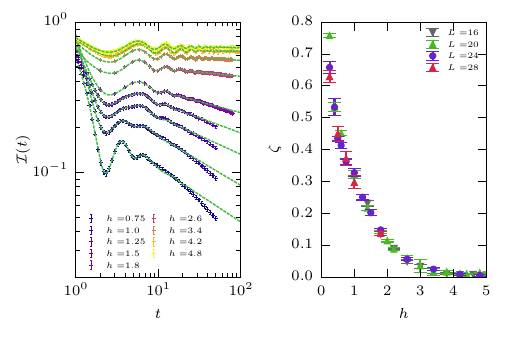}
    \caption{Left: Time evolution of the disorder averaged spin imbalance for a chain of length $L=24$. Lines are best fits to Eq.
    \eqref{eq:imbalance_fit}.  Right: Exponent $\zeta$ of the spin imbalance
    decay as a function of disorder strength $h$, as extracted from fits for different system sizes.
All systems have periodic boundary conditions. The results for $h > h_c \simeq 3.7$ are compatible
with $\zeta=0$ up to systematic and statistical errors. }
    \label{fig:imbalance}
\end{figure}

Here, we show that the intermediate time dynamics of the imbalance can display an anomalous
power-law regime characteristic of the sub-diffusive regime. We generalize the imbalance to any
initial basis state of the form $\ket{\psi(0)} =\ket{\sigma_1,\dots,\sigma_L}$ (with zero
magnetization) presenting a trivial local spin imbalance, by computing
\begin{equation} \mathcal{I}(t) =
    \frac{4}{L} \sum_{j=1}^{L}\bra{\psi(0)}S_j^z(0) S_j^z(t)\ket{\psi(0)}, \label{eq:imbalance_def}
\end{equation} 
for $L$ (even) sites.  Shown in panels c) and d) of Fig.~\ref{fig:overview} and in Fig.~\ref{fig:imbalance}, 
the disorder-averaged imbalance ${\cal{I}}(t)$ displays as expected qualitatively
different behaviors for ETH and MBL regimes. Below we focus on the delocalized side where the imbalance is vanishing at long time.

There, an anomalous power-law regime with varying exponents is found at intermediate time (Fig.~\ref{fig:imbalance}), even if hindered by strong and fast oscillations at short time $t \lesssim 10$. This transient behavior, particularly pronounced at small disorder, is reminiscent of the clean case where these oscillations are exponentially suppressed in time~\cite{barmettler_relaxation_2009,barmettler_quantum_2010}. We find that the best fitting function faithfully describing the entire relaxation at intermediate times is given by
\begin{equation}
\mathcal{I}(t)=a{\rm{e}}^{-\frac{t}{\tau}}\cos(\omega_1 t + \varphi)+bt^{-\zeta}[1+ct^{-\eta}\sin(\omega_2
t+\varphi)].
\label{eq:imbalance_fit}
\end{equation}
The first term is identical to the clean
case~\cite{barmettler_relaxation_2009,barmettler_quantum_2010}, while the second contains the
anomalous power-law characterized by the exponent $\zeta$. The final oscillatory term with the
subdominant power-law ($\eta>0$) describes the characteristic oscillations that are visible inside
the power-law regime, and which is found to be out-of-phase with the first term. The dashed lines in
Fig.~\ref{fig:imbalance} represent fits to this form, and an excellent agreement with the raw data
(symbols) can be observed. Note that this fitting form does not capture the finite-size saturation at longer times in the ETH regime, which is visible at low disorder in our time regime (fit windows are chosen accordingly to exclude this finite-size effect in this region). The extracted exponent $\zeta$ (Fig.~\ref{fig:imbalance}b) vanishes at
the MBL transition, and monotonously increases when disorder is reduced. This is confirmed by the
good agreement obtained between exponents extracted for systems of different sizes. For weak
disorder strength, the extraction of $\zeta$ is more difficult for two concomitant
reasons: (i) the short-time exponential oscillatory decay is very strong and has already depleted
strongly the imbalance, leaving only a small time window to observe the power-law regime, which is
furthermore cut by (ii) a saturation to a non-zero long-time value of the imbalance for a
finite-size system, visible as strong size dependence of the result for $h\lesssim0.5$, limiting
the reliability of our result at very small disorder strength. 

Finally in the localized phase, the fit to Eq.~\ref{eq:imbalance_fit} is particularly
good up to very long times as $\zeta$ is found to vanish, leaving a finite long-time saturation value for the imbalance (the exact vanishing of $\zeta$ within error bars requires to consider longer times on smaller
systems to observe saturation). In the MBL regime, we observe that $\omega_2\approx1$ and $\eta$
decreases slowly with disorder strength, starting from $\eta\approx 1$ at the transition, fully
consistent with the expectation that the oscillations around the saturation value decay as a power
law~\cite{serbyn_quantum_2014}.\\
{\it Discussion---} Our large-scale exact numerical results confirm the existence of an anomalous
dynamical regime for the entanglement entropy inside the ETH phase, as predicted in
Refs.~\onlinecite{vosk_theory_2015,potter_universal_2015}. This slow dynamical behavior can be
probed in cold-atom systems~\cite{schreiber_observation_2015} by measuring the power-law decay of
imbalance at intermediate time as we have clearly shown.  While it is hard to conclude on the
behavior at very small disorder, we strikingly find that this anomalous regime persists in an extended parameter region for a large window of disorder. At first sight, this may be hard to reconcile with the fact that the sub-diffusive regime is
ascribed~\cite{vosk_theory_2015,potter_universal_2015,agarwal_anomalous_2015} to rare Griffiths
regions, only expected close to the MBL transition. One should remember however that in the
considered quench protocol, inhomogeneity is also present in the initial random product state $|
\psi(0)\rangle$, where the energy density can fluctuate locally leading to anomalously hot or cold
regions. We believe that the presence of a mobility edge~\cite{luitz_many-body_2015} in the model
Eq.~\eqref{eq:Hrf} can therefore enhance the extent of the anomalous dynamical regime when using
such a global quench protocol. It would be interesting in future work to consider the full
range of energy for the initial state to see whether dynamics can also detect the presence of a
mobility edge. Also, the subdiffusive regime is expected to occur
only in one dimension~\cite{agarwal_anomalous_2015}: while it is a challenging task to extend
numerical simulations of MBL to two-dimensional systems, it is possible with our numerical technique
to address the dynamical behavior of a ladder geometry where a MBL phase has been recently
found~\cite{baygan_many-body_2015}. We leave these interesting questions to future studies.
\begin{acknowledgments}
We thank E. Altman, B. Clark, S. Parameswaran, D. Poilblanc and R. Vasseur for very useful discussions. 
This work was supported in part by the Gordon and Betty Moore Foundation's EPiQS Initiative through Grant No. GBMF4305 at the University of Illinois and the French ANR program ANR-11-IS04-005-01. 
Our code is based on the PETSc~\cite{petsc-web-page,petsc-user-ref,petsc-efficient} and
SLEPc~\cite{hernandez_slepc:_2005} libraries. This work was performed using HPC
resources from GENCI (grant x2015050225) and CALMIP (grant 2015-P0677). 
\end{acknowledgments}

\bibliography{mbl_quench,mbl_quench_manual}

\appendix

\end{document}